\documentclass[aps,showpacs,twocolumn,showkeys,prl]{revtex4}
\usepackage{graphicx,color,epsfig,rotate}
\usepackage{times,xspace}
\usepackage{amsbsy,amssymb,amsmath,bm}
\usepackage{dcolumn}
\usepackage{bm}
\input{epsf.sty}

\def\bbbc{{\mathchoice {\setbox0=\hbox{$\displaystyle\rm C$}\hbox{\hbox
to0pt{\kern0.4\wd0\vrule height0.9\ht0\hss}\box0}}
{\setbox0=\hbox{$\textstyle\rm C$}\hbox{\hbox
to0pt{\kern0.4\wd0\vrule height0.9\ht0\hss}\box0}}
{\setbox0=\hbox{$\scriptstyle\rm C$}\hbox{\hbox
to0pt{\kern0.4\wd0\vrule height0.9\ht0\hss}\box0}}
{\setbox0=\hbox{$\scriptscriptstyle\rm C$}\hbox{\hbox
to0pt{\kern0.4\wd0\vrule height0.9\ht0\hss}\box0}}}}

\begin{document}
\title{Magnetoelectric effects in an organo-metallic quantum magnet}

\author{V. S. Zapf$^1$, F. Nasreen$^{1,2}$, F. Wolff-Fabris$^{1,3}$ and A. Paduan-Filho$^4$}

\affiliation{
$^1$National High Magnetic Field Laboratory (NHMFL), MPA-CMMS, Los Alamos National Lab, Los Alamos, NM \\
$^2$New Mexico State University, Las Cruces, NM \\
$^3$Now at Dresden Hochfeld Labor, Dresden, Germany \\
$^4$Instituto de Fisica, Universidade de Sao Paulo, Brazil \\
}

\date{\today}

\begin{abstract}
We observe a bilinear magnetic field-induced electric polarization of 50 $\mu C/m^2$ in single crystals of NiCl$_2$-4SC(NH$_2$)$_2$ (DTN).  DTN forms a tetragonal structure that breaks inversion symmetry, with the highly polar thiourea molecules all tilted in the same direction along the c-axis. Application of a magnetic field between 2 and 12 T induces canted antiferromagnetism of the Ni spins and the resulting magnetization closely tracks the electric polarization. We speculate that the Ni magnetic forces acting on the soft organic lattice can create significant distortions and modify the angles of the thiourea molecules, thereby creating a magnetoelectric effect. This is an example of how magnetoelectric effects can be constructed in organo-metallic single crystals by combining magnetic ions with electrically polar organic elements.
\end{abstract}

\maketitle

\epsfxsize=250pt
\begin{figure}[tbp]
\epsfbox{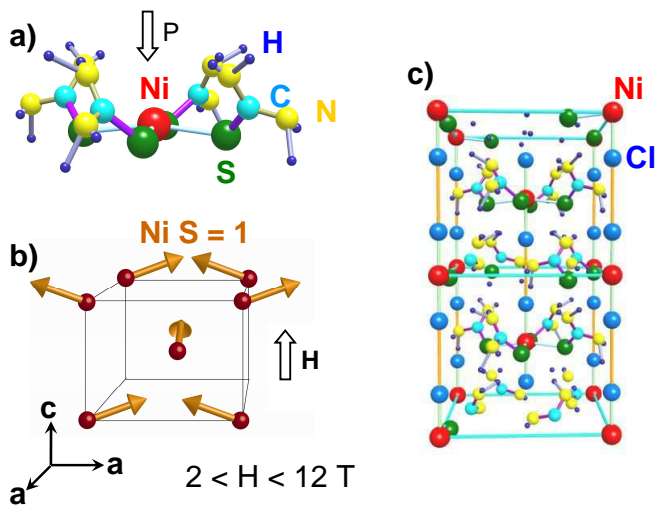}
\caption{Crystal structure of NiCl$_2$-4SC(NH$_2$)$_2$ (DTN) showing a) the thiourea coordination around the Ni ions and the likely direction of the electric polarization $P$, b) the canted frustrated antiferromagnetic structure of the Ni spins at intermediate fields between 2 and 12 T, and c) two unit cells of the full crystal structure.}
\label{DTN_structure}
\end{figure}

Magnetoelectrics are compounds in which the magnetic and electric susceptibilities are coupled, with magnetic fields inducing electric polarization and electric fields inducing magnetic polarization \cite{Fiebig05}. Research in this field is motivated by the promise of new devices as well as improving the speed, energy-efficiency and size of existing circuits \cite{Cheong07,Scott07,Chu08}. The magnetoelectric effect can be particularly large when either the magnetic or electric subsystem is ordered, leading to diverging magnetic or electric susceptibilities. In particular, multiferroic behavior is a current hot topic that attempts to exploit the large effects that can result when both the magnetic and electric polarizations in electric insulators exhibit long-range order \cite{Kimura03,Goto04,Hur04,Lawes05,Katsura05,Cheong07,Kenzelmann07,Arima07,Kimura07}. However, there is currently a dearth of materials exhibiting strong magnetoelectric coupling or multiferroic behavior and most research to date in these fields, particularly in multiferroics, has focused on transition-metal oxides. Thus the idea of using organic ferroelectrics as a starting point for building multiferroic or magnetoelectric materials has the potential to greatly expand the available number of compounds \cite{Zapf10,Kagawa10}. Ferroelectricity is known to occur in a number of organic molecules \cite{Horiuchi08}, and was most recently discovered above room temperature in croconic acid \cite{Horiuchi10}. Organic materials tend to possess soft lattice structures that can be easily modified by magnetic forces, leading to large magnetic field-induced changes in the electric polarization. In addition, the flexibility for design of organic molecules and availability of electrically polar building blocks could allow us to construct a variety of new magnetoelectrics by design. A recent review of ferroelectricity in organic materials \cite{Horiuchi08} has identified the electrically polar molecule thiourea, SC(NH$_2$)$_2$, as a strong candidate for organic ferroelectricity. In its crystalline form it is a ferroelectric with a $T_c$ of 169 K and an electric polarization of 3,200 $\mu C/m^2$ \cite{Goldsmith59}.  The origin of the electric polarization is primarily the polar double bond between carbon and the highly electronegative sulfur atom. This bond is only partially compensated by the remainder of the molecule. In pure thiourea, the individual thiourea molecules nearly but not quite anti-align, and an applied electric field can tilt the relative orientations, resulting in a ferroelectric response. 

\begin{figure}[tbp]
\epsfbox{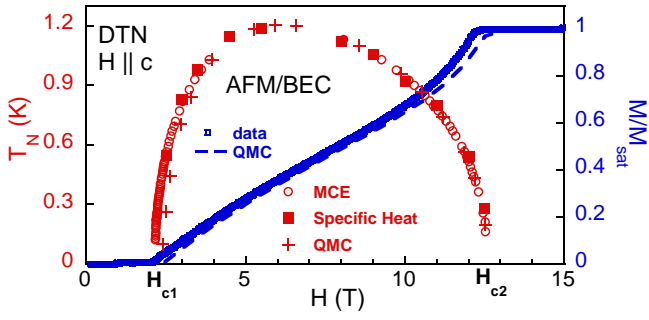}
\caption{Temperature $T$ - Magnetic field $H$ phase diagram for $H || c$ determined from
specific heat and magnetocaloric effect (MCE) data, together with
the result of Quantum Monte Carlo (QMC) simulations. The magnetization vs field measured at 16
mK and calculated from QMC is overlaid onto the phase diagram. The region of antiferromagnetism/Bose-Einstein Condensation (AFM/BEC) occurs under the red dome. \cite{Zapf08}}
\label{DTN_phase_diagram}
\end{figure}

Here we present data on the compound NiCl$_2$-4SC(NH$_2$)$_2$ (DTN) in which the thiourea molecules all tilt in the same direction along the tetragonal c-axis, thereby breaking spatial inversion symmetry (see fig. \ref{DTN_structure}a).  The Ni magnetic spins occupy a tetragonal body-centered structure with antiferromagnetic interactions between nearest neighbors. Frustration between the two interpenetrating tetragonal sublattices of Ni spins causes the spins of the sublattices orient orthogonal to each other (see fig. \ref{DTN_structure}b). This compound was previously investigated for its quantum magnetism that can be modeled in terms of Bose-Einstein Condensation of the spin system
\cite{PaduanFilho04,Zapf06,Zvyagin07,Zapf08,Zvyagin08,Yin08,Cox08,Chiatti09,Zherlitsyn09,PaduanFilho09} and thus extensive information about its magnetic and magneto-elastic properties is available. In zero magnetic field, a uniaxial anisotropy $D \sim 9$~K of the Ni spins splits the Ni $S$ = 1 triplet into a $S_z = 0$ ground state and
a $S_z = \pm 1$ excited doublet. The $S_z = 1$ state can be suppressed with applied magnetic fields along the tetragonal
\textsl{c}-axis via the Zeeman effect until it becomes degenerate with the $S_z = 0$ state, thus producing a magnetic ground state above $H_{c1} = 2.1$~T. As shown in Fig. \ref{DTN_phase_diagram}, antiferromagnetic super-exchange between the Ni atoms produces long-range antiferromagnetic order in a dome-shaped region of the $T-H$ phase diagram between $H_{c1} = 2.1$ T, where the magnetic ground state is induced, and $H_{c2} \sim 12$ T where the spins align with the applied magnetic field \cite{Zapf06}. The largest antiferromagnetic exchange occurs along the c-axis with $J_c = 1.8$~K, resulting in an effective energy scale of $\sim 10$ K below which the longitudinal magnetization along the c-axis becomes significant. 3-D long-range order occurs below 1 K due to the weaker antiferromagnetic exchange along the a-axis with $J_a = 0.18$ K. The frustrated diagonal coupling is estimated to be less than 20 mK \cite{Zvyagin08}. Spin-lattice coupling between the magnetic spins and the lattice has been observed via magnetostriction \cite{Zapf08} and sound velocity \cite{Chiatti09,Zherlitsyn09}. In effect, the Ni atoms attract each other along the c-axis when they are anti-aligned and their antiferromagnetic bonds are satisfied and repel each other when they are aligned, with more complicated behavior occurring along the a-axis. The bulk modulus measured using resonant ultrasound spectroscopy is $E_{33} = 7.5 \pm 0.7$ GPa extrapolated to zero K \cite{Zapf08}, which is an order of magnitude smaller than typical inorganic metals and oxides.

\epsfxsize=250pt

\epsfxsize=200pt
\begin{figure}[tbp]
\epsfbox{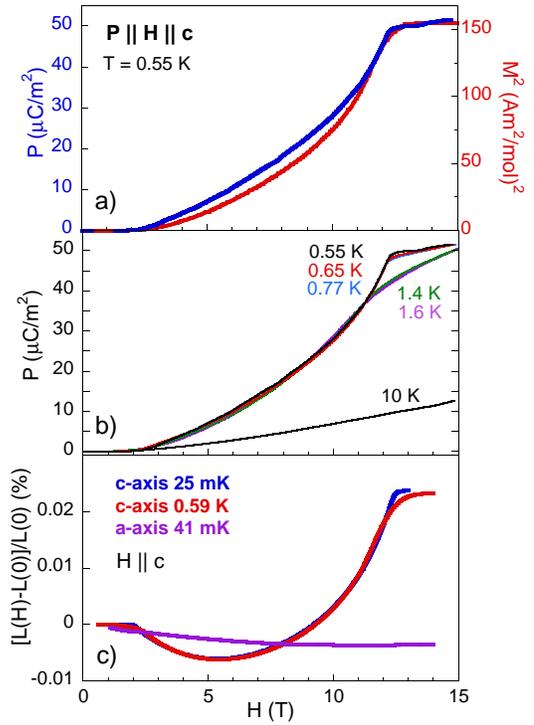}
\caption{a) Comparison of the electric polarization $P$ (this work) and the square of the magnetization $M^2$ vs $H$ \cite{PaduanFilho04} at 0.6 K. b) $P(H)$ of DTN at various temperatures. c) Percentage change in length $L$ of the sample as a function of magnetic field relative to its zero field value for $ H || c$, measured along the crystallographic a and c axes \cite{Zapf08}. }
\label{DTN_comparison}
\end{figure}

Single crystals of DTN were grown from aqueous solutions of thiourea and nickel chloride. Two sets of samples were measured, one grown in an electric field of 500 V/cm along the c-axis and the other in zero electric field. Comparable electric polarizations were observed to within 25\% and data on the E-field grown samples are presented.  The electric polarization as a function of magnetic field was measured at temperatures down to 0.55 K with the samples immersed in pumped $^3$He. Pulsed magnetic fields were employed at the National High Magnetic Field Laboratory (NHMFL) at Los Alamos National Laboratory. A relatively slow mid-pulse magnet ($\sim 2000$ T/s peak, 500 T/s average) was used to avoid heating and cooling from eddy currents and magnetoelectric effects. Capacitor plates were constructed by applying Dupont silver paint to the faces of the single crystals perpendicular to the c-axis, and the electric polarization along the c-axis for magnetic fields along the c-axis was measured with a Stanford Research 560 current-to-voltage amplifier.
Dielectric constant measurements as a function of magnetic field were also performed capacitatively using a GC capacitance bridge driven at 30 V and 5 kHz with $H$ and $E$ along the c-axis. However no magnetic field-dependence could be observed of the dielectric constant.

The electric polarization as a function of magnetic field $P(H)$ is shown in Fig. \ref{DTN_comparison}a measured during the up and down sweeps of a pulsed magnet with $P$ and $H$ along the crystallographic c-axis at 0.55 K. The square of the magnetization $M^2(H)$ is shown for comparison\cite{PaduanFilho04} since it not only tracks $P(H)$ more closely than $M$ but, like $P$ is also even under reversal of the magnetic field. The $P(H)$ data was taken after cooling the sample in zero electric and magnetic field, e.g. with no poling. The electric polarization and the square of the magnetization show a similar magnetic field dependence, being roughly zero for fields up to $H_{c1} \sim 2$ T, and then increasing strongly in the canted antiferromagnetic state until saturization occurs above $H_{c2} \sim 12 $T.  In Fig. \ref{DTN_comparison}b, $P(H)$ curves are shown for different temperatures up to 10 K. The magnitude of the jump in $P$ with $H$ remains roughly constant up to 1.6 K, even though this exceeds the maximum antiferromagnetic ordering temperature and only becomes suppressed in the 10 K curve. This behavior is also similar to $M(H)$, as shown in ref. \cite{PaduanFilho04}, where the magnitude of $M(H)$ is primarily determined by the $\sim 10$ K energy scale of the largest antiferromagnetic interaction $J_c$. The signature of long-range ordering is by contrast relatively small. In the $P(H)$ curves, a kink due to antiferromagnetic ordering was also very small and could only be distinguished for some of the curves.

The magnetostriction $(L(H)-L(0))/L(0)$ \cite{Zapf08} is shown in Fig. \ref{DTN_comparison}c, measured for magnetic fields along the c-axis, and for length changes along both a and c. The magnetostriction shows clear signatures of the magnetic behavior of the system, particularly along c. However, the behavior is clearly quite complicated with non-monotonic behavior along c and monotonic behavior along a. The magnetostriction for $L || c$ curve at 25 mK was well fit by a 1-D theory in ref. \cite{Zapf08}, which assumes that the Ni ions attract and repel each other along the c-axis depending on their relative spin orientations, and assuming that these forces are caused by a to first-order linear dependence of $J_c$ on bond length acting in competition with the elastic energy of the lattice. We posit that the thiourea molecules are the dominant source of electric polarization in DTN, since these contain highly polar S=C bonds that can cause electric polarizations up to 3,200 $\mu C/m^2$ in pure thiourea. The exact relationship between the angles of these thiourea molecules and the overall length changes measured using magnetostriction is not easily calculated given the available data, but is likely a function of both the nonmonotonic behavior along the c-axis and the monotonic behavior along the a-axis. 

In conclusion, we observe a magnetoelectric effect of 50 $\mu C/m^2$ in the organo-metallic compound NiCl$_2$-4SC(NH$_2$)$_2$ that is bilinear (e.g. even under reversal of the magnetic field) and likely results from magnetic forces between Ni spins distorting the soft organic lattice. We propose that the polar thiourea molecules, in particular the double carbon-sulfur bonds, are the likely origin of the electric polarization, and their angles are modified due to magnetostrictive effects in this compound. This compound is magnetoelectric -- we do not see any evidence of ferroelectric behavior that would render this a multiferroic compound. However, the size of the magnetic field-induced electric polarization, 50 $\mu C/m^2$ is only an order of magnitude smaller than some commonly studied multiferroics, e.g. TbMnO$_3$ with 600 $\mu C/m^2$) \cite{Kimura07}.
This is the first example of the organic ferroelectric thiourea being used as a building block to create magnetoelectric or multiferroic behavior. Given the availability of thiourea and other organic ferroelectrics with relatively large ferroelectric polarizations and high ordering temperatures, coupled with the soft lattice structures of organo-metallic compounds, we propose this hybrid approach as an interesting area for further growth. 

Very recently multiferroic behavior has also been reported in a few other organo-metallics materials including CuCl$_2$-2[(CH$_3$)$_2$SO] \cite{Zapf10} and TTF-BA \cite{Kagawa10}. The combination of stronger magneto-electric couplings and higher temperatures of the magnetic ordering are now a necessary step to further develop this field.

\begin{acknowledgements}

Work at the National High Magnetic Field Laboratory was supported by the U.S. National Science Foundation through Cooperative Grant No. DMR901624, the State of Florida, and the U.S. Department of Energy. A.P.F. acknowledges support from CNPq (Conselho Nacional de Desenvolvimento Científico e Tecnológico, Brazil).

\end{acknowledgements}


\end{document}